\documentclass[prb,twocolumn]{revtex4-1}
\usepackage{amsmath}
\usepackage{amsfonts}
\usepackage{graphicx}

\begin{document}
\title{Is the tautochrone curve unique?}
	
\author{Pedro Terra}
\email{terra@if.ufrj.br}
\author{Reinaldo de Melo e Souza}
\email{reinaldo@if.ufrj.br}
\author{C. Farina}
\email{farina@if.ufrj.br}

\affiliation{Universidade Federal do Rio de Janeiro,
Instituto de F\'isica, Rio de Janeiro, RJ 21945-970}

\date{\today}

\begin{abstract}

The answer to this question is no. In fact, in addition to the solution first obtained by Christiaan Huygens in 1658, given by the cycloid, we show that there is an infinite number of tautochrone curves.
With this goal, we start by briefly reviewing an the problem of finding out the possible potential energies that lead to periodic motions of a particle whose period is a given function of its mechanical energy.
There are infinitely many solutions, called sheared potentials. As an interesting example, we show that a P\"oschl-Teller and the one-dimensional Morse potentials are sheared relative to one another for negative energies, clarifying why they share the same periods of oscillations for their bounded solutions. We then consider periodic motions of a particle sliding without friction over a track around its minimum under the influence of an uniform gravitational field. After a brief historical survey of the taucochrone problem we show that, given the period of oscillations, there is an infinity of tracks which lead to the same periods. As a bonus, we show that there are infinitely many tautochrones.

\end{abstract}

\maketitle

\section{Introduction}

In classical mechanics we find essentially two kinds of problems: {\it (i)} the fundamental (or direct) problems, in which given the forces on a system we must obtain its possible motions, and {\it (ii)} inverse problems, in which we know the possible motions of a given system (or some characteristics thereof), and must determine the forces that caused them. In an informal way, we could say that in inverse problems we attempt to find the causes by analyzing the effects. A humorous manner of stating the difference between these two kinds of problems, which can be found in Bohren and Huffmann's book \cite{BH}, is to say that in the direct problem, given a dragon we should be able to determine its tracks, while in the inverse problem, from the simple observation of the dragon's track we should be able to infer what the dragon is like. 

Some historically relevant examples of inverse problems are Newton's determination of the gravitational force from Kepler's laws, and Rutherford's discovery of the atomic nucleus from the scattering of $\alpha$ particles from a very thin sheet of gold. 
Inverse problems are, indeed, extremely frequent even in contemporary physics. 
In high-energy Physics, for instance, we try to understand the fundamental interactions between elementary particles by observing the products of their scattering. Oil prospecting methods in deep waters are also inverse problems as they rely on analyzing the properties of the reflected waves (caused by small explosions at the surface of the sea) by the presence of interfaces separating two mediums within the surface of the Earth.

Even in the case of a particle moving in one dimension under the influence of a conservative total force, analyzing inverse problems may yield very surprising results. For instance, consider the periodic motions of a point mass in a potential well $U(x)$. Although the knowledge of $U(x)$ uniquely 
determines the periods of oscillations as a function of the mechancal energy $\tau(E)$, the opposite is not true, since the knowledge of $\tau(E)$ does not uniquely determine the potential energy that caused those periods. In fact, as explained in the next section, it can be shown that one given function $\tau(E)$ allows for infinitely many potential wells which are sheared from one another. 

This beautiful result was obtained in a quite ingenious way by Landau and Lifshitz \cite{Landau-mechanics}. Since then, it has been revisited by some authors using different approaches. For instance, Pippard's pleasant book \cite{Pippard} presents a neat graphical demonstration of sheared potentials, whereas Osypowski and Olsson \cite{Osypowski-1986} provide a different demonstration with the aid of Laplace transforms. For recent developments on this topic and analogous quantum cases, see Asorey {\it et al.}\cite{Asorey-2007}.

In this article, our main goal is to analyze inverse problems for periodic motions of a particle sliding on a frictionless track contained in a vertical plane in a uniform gravitational field. 
Similarly to the sheared potentials, the knowledge of the shape of the track determines uniquely the period of motion as a function of the maximum height achieved by the particle. 
However, as we will demonstrate, an infinity of different curves provide the same periods for every maximum height. 
We will show that these curves exhibit a geometrical property akin to shearing which, to our knowledge, had not been noted previously in the literature.

In particular, we are interested in the case in which the period of oscillations does not depend on the maximum height (and, therefore, on the mechanical energy) of the system, that is, the tautochrone. Since this term is sometimes associated to a downward-only motion, we employ the expression {\it round-trip tautochrone} to specify our curves of interest. It has been shown by Huygens in 1658 that the cycloid is a tautochrone curve. A direct consequence of our analysis is that this is not the sole solution of the tautochrone problem. In fact, there are infinitely many round-trip tautochrone tracks, unfolding new results for this three-century-old problem.

The article is organized as follows: In section \ref{sec:shearedreview}, we review the basic concepts of sheared potentials following Landau and Lifshitz\cite{Landau-mechanics}. We also provide two non-trivial examples of sheared potentials, namely: {\it (i)} power-law potentials and {\it (ii)} a P\"oschl-Teller and the one-dimensional Morse potentials. In section \ref{sec:generalization}, inspired by the properties exhibited by sheared potential wells presented in section \ref{sec:shearedreview}, we consider periodic motions of a particle sliding without friction and under a constant gravity on a track contained in a vertical plane, and ask what curves yield the same periods of oscillations. We show that, in truth, there is an infinite number of different curves associated to the same periods. We then give a nice interpretation for the condition between these curves as a new kind of shearing. In section \ref{sec:tautochrone} we obtain, from the previous result, an infinite family of tautochrone curves. Section \ref{sec:remarks} is left for final remarks.

\section{Brief review of sheared potential wells}
\label{sec:shearedreview}

Let us consider a particle of mass $m$ moving along the ${\cal O}x$ axis under the influence of the resultant force ${\cal F}(x) = -dU(x)/dx$, where $U(x)$ is a generic potential well with a single minimum. Without loss of generality, let us choose the origin of the ${\cal O}x$ axis at the minimum of the potential well such that $U(0) = 0$. Suppose the particle moves with mechanical energy $E$. The associated turning points are simply given by the roots of the algebraic equation $E = U(x)$, which we denote by $x_1$ and $x_2$. It is worth emphasizing that $x_1$ and $x_2$ depend on $E$. Assuming $x_2 > x_1$, it is straightforward to use the conservation law of the mechanical energy to obtain the period of oscillations:
\begin{equation}
\label{eq:periodo-pot}
\tau(E)=\sqrt{2m}\int_{x_1}^{x_2}\frac{dx}{\sqrt{E-U(x)}} \, .
\end{equation}
It is evident from the previous equation that the potential energy $U(x)$ and a given mechanical energy $E$ uniquely determine the period of the oscillations. In other words, the function $\tau: E\;\longrightarrow\;\tau(E)$ is uniquely determined by the knowledge of function $U:x\;\longrightarrow\; U(x)$.

In order to solve the inverse problem, namely, given $\tau(E)$ to find the corresponding $U(x)$, it is tempting (and convenient) to regard the coordinate $x$ as a function of the potential $U$. However, since the function $U(x)$ is not injective, obtaining its inverse requires defining two functions, namely: $x_L: U \longmapsto x_L(U)$, for the left branch of $U(x)$ ($x<0$) and $x_R: U \longmapsto x_R(U)$, for the right branch of $U(x)$ ($x\geq 0$). Performing a change of variables, we recast Eq. (\ref{eq:periodo-pot}) into the form
\begin{align}
\label{eq:periodoLR}
\tau(E)&=\sqrt{2m} \left[ \int_{0}^{E}\frac{dx_R}{dU}\frac{dU}{\sqrt{E-U}} + \int_{E}^{0}\frac{dx_L}{dU}\frac{dU}{\sqrt{E-U}} \right] \nonumber \\
&= \sqrt{2m} \int_{0}^{E}\left(\frac{dx_R}{dU}-\frac{dx_L}{dU}\right)\frac{dU}{\sqrt{E-U}} \, .
\end{align}
To get rid of the integral on the right-hand side, we multiply both sides of Eq. (\ref{eq:periodoLR}) by $dE/\sqrt{\alpha-E}$, where $\alpha$ is a constant parameter, 
and then integrate from $0$ to $\alpha$. On the right-hand side, we are left, then, with an integral in $U$ followed by an integral in $E$. Changing the order of integration, which requires a subtle change in the integration limits in order to preserve the integration region in the $E-U$ plane, we obtain

\begin{eqnarray}
\label{eq:almostthere}
\int_0^\alpha \frac{\tau(E)dE}{\sqrt{\alpha-E}} &=& 
 \sqrt{2m}\int_0^\alpha \left[\frac{dx_R}{dU}-\frac{dx_L}{dU}\right]dU \times\cr\cr
 &\times& 
 \int_U^\alpha\frac{dE}{\sqrt{\alpha-E}\sqrt{E-U}} \, .
\end{eqnarray}
Fortunately, the integral over $E$ may be exactly calculated and does not depend on $U$. In fact, it is straightforward to show that this integral is 
equal to $\pi$ (it can be computed, for instance, by completing the square). As a consequence, the integration on $U$ is immediate. From these results, and rewriting $\alpha=U$, 
we conclude that a given period function $\tau$ does not determine a single potential $U$, but rather an infinite family of potentials satisfying the relation
\begin{equation}
\label{eq:xRxL}
x_R(U)-x_L(U)=\frac{1}{\pi\sqrt{2m}}\int_0^U\frac{\tau(E)dE}{\sqrt{U-E}} \, .
\end{equation}

The last equation means that, provided two potential wells $U(x)$ and $\tilde{U}(x)$ share the same
width $x_R(U)-x_L(U)=\tilde{x}_R(U)-\tilde{x}_L(U)$ for every value of $U$, the periods of the oscillations of a particle under the influence of these potentials will have periods described by the same function 
$\tau(E)$. When this property is satisfied, we say that the potential wells $U(x)$ and $\tilde U(x)$ are sheared relative to one another. However, if we further require the potential to be symmetric, that is $x_R(U)=-x_L(U)$, then it is uniquely determined by the period function $\tau$. 

The aforedescribed development was reproduced from Landau and Lifshitz\cite{Landau-mechanics}. Let us illustrate the previous results in some non-trivial examples that are not usually dealt with in the literature. We will start by illustrating the shearing process for the power-law case. Then, we will show that the Morse and the P\"osch-Teller potentials are related by shearing (for bounded motions).

\subsection{Power law potentials}
\label{ssec:powerlaw}

Sheared potentials derived from the parabolic potential well have been considered by Ant\'on and Brun \cite{Anton-parabolashear2008} . These authors obtained anharmonic motions whose periods of oscillations retain the property of being independent from the energy (isochronous oscillations).

In this subsection, we shall consider the more general case of the family of symmetric power law potentials given by $U(x) = a |x|^\nu$, with $a>0$ and $\nu\ge 1$. The turning points are $x_{\pm}=\pm(E/a)^{1/\nu}$. The period of oscillations may be computed by use of Eq. (\ref{eq:periodo-pot}):
\begin{align}
\label{eq:periodo_powerlaw}
\tau(E) &= \sqrt{2m}\int_{x_-}^{x_+} \frac{dx}{\sqrt{E - a|x|^\nu}} \nonumber \\
&= x_+^{-\nu/2} \sqrt{\frac{8m}{a}} \int_0^{x_+}\left\{1-\left(\frac{x}{x_+}\right)^\nu\right\}^{-1/2}dx \, ,
\end{align}
where we used the fact that the potential well is symmetric and that $E = a x_+^\nu$. Making a change of variables $u=(x/x_+)^\nu$ and recalling the definition for $x_+$ we rewrite the period as
\begin{equation}
	\label{eq:periodo_powerlaw2}
	\tau(E) = E^{\left(\frac{1}{\nu}-\frac{1}{2}\right)} \frac{\sqrt{8m}}{a^{1/\nu}} \frac{I(\nu)}{\nu}\, ,
\end{equation}
where $I(\nu)=\int_0^1 u^{1/\nu - 1}(1-u)^{-1/2}du$ is simply a numerical factor. Although the computation may be continued to give an exact result\cite{Landau-mechanics,CarinenaFarinaSigaud-93}, it suffices for the purpose of this article to regard the period's dependence on the energy. Moreover, Eq. (\ref{eq:periodo_powerlaw2}) shows that for $\nu=2$ the exponent on $E$ vanishes and, as a consequence, the period is independent of the energy, which is the case of the harmonic oscillator.

Let us denote by $D$ the distance between the turning points for a given energy $E$, that is, $D(E)=x_+(E)-x_-(E)$. We may try to construct a sheared 
potential for $U(x) = a\vert x\vert^\nu$, denoted by $\tilde U(x)$, by defining different expressions for positive and negative $x$:
\begin{equation}
\tilde{U}(x) = \left\{ \begin{array}{lr} b|x|^\nu & : x < 0 \\ 
c|x|^\nu & : x \ge 0 \end{array} \right.
\end{equation}
whose difference between the two turning points corresponding to energy $E$ is given by 
\begin{equation}
\tilde{D}(E) = \tilde x_+(E) - \tilde x_-(E) = (E/b)^{1/\nu}+(E/c)^{1/\nu}\, .
\end{equation}
Imposing the shearing condition, namely, $D(E) = \tilde D(E)$, for any $E$, we obtain the following condition between coefficients $a$, $b$ and $c$:
\begin{equation}
\label{eq:coef_powerlaw}
\frac{1}{a^{1/\nu}} = \frac{1}{2}\frac{1}{b^{1/\nu}} + \frac{1}{2}\frac{1}{c^{1/\nu}} \, .
\end{equation}
Indeed, if we compute explicitly the period of the oscillations of the particle under the influence of the potential well $\tilde U(x)$ by suitably adapting Eq. (\ref{eq:periodo_powerlaw2}) we shall obtain
\begin{equation}
	\label{eq:periodo-powerlaw-shear}
	\tilde{\tau}(E) = \left[\frac{1}{2}\frac{1}{b^{1/\nu}}+\frac{1}{2}\frac{1}{c^{1/\nu}} \right] E^{\left(\frac{1}{\nu}-\frac{1}{2}\right)} \sqrt{8m} \frac{I(\nu)}{\nu} \, .
\end{equation}
Substituting Eq. (\ref{eq:coef_powerlaw}) into Eq. (\ref{eq:periodo-powerlaw-shear}) it becomes evident that $\tilde\tau(E) = \tau(E)$, as expected, since we forced potential wells $U(x)$ and $\tilde U(x)$ to be sheared relative to each other.

\subsection{Morse and P\"oschl-Teller potentials}

\begin{figure}[h!]
	\begin{center}
		\includegraphics[width=\linewidth]{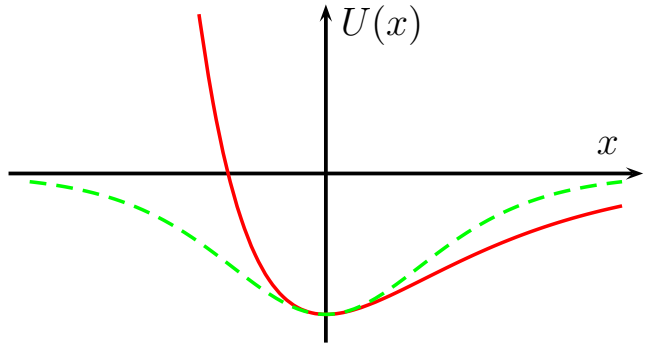}
		\caption{Morse (solid red line) and P\"oschl-Teller (dashed green line) potentials.}
		\label{fig:morsePT}
	\end{center}
\end{figure}

Let us now consider two less trivial potentials, namely, the one-dimensional Morse potential and the P\"oschl-Teller potential, denoted by $U_M(x)$ and $U_{PT}(x)$, respectively, and given by (see Fig. \ref{fig:morsePT})
\begin{eqnarray}\label{Morse}
U_M(x) &=& U_0 \left(e^{-2\alpha x}-2e^{-\alpha x}\right)\\
U_{PT}(x) &=& - \, \frac{U_0}{\cosh^2(\alpha x)}\, ,
\label{PT}
\end{eqnarray}

where $U_0$ and $\alpha$ are positive constants. These two potentials have their importance also beyond the context of classical mechanics. For instance, the Morse potential plays an important role in chemistry while the P\"oschl-Teller potential exhibits bizarre properties when treated quantum mechanically. To mention just one of them, for appropriate values of the constant $U_0$ this potential becomes reflectionless for whatever value of the energy of the particle \cite{Lekner}.

Our purpose in this subsection is to show that these two potentials are sheared relative to each other, so that they have the same period of oscillations for negative mechanical energies. 
With this goal, let us start by determining the turning points of a particle moving in the Morse potential with a negative mechanical energy $E$. These points can be obtained by solving equation $E=U_M(x)$. Conveniently substituting $u=e^{-\alpha x}$, this equation leads to
\begin{equation}
\label{eq:changevarMorse}
u^2 - 2u - \frac{E}{U_0} = 0\, ,
\end{equation}
whose two roots are given by
\begin{equation}
u_{\pm}=1\pm\sqrt{1 + E/U_0}\, .
\end{equation}
Getting back to the variable $x$, we have
\begin{equation}
\label{eq:turningMorse}
x_{\pm}=\frac{1}{\alpha}\ln(1\pm\sqrt{1+E/U_0})\, .
\end{equation}
Hence, the distance between these two turning points, $D_M(E) = x_+(E) - x_-(E)$, for an arbitrary negative energy is given by
\begin{eqnarray}
\label{eq:diamMorse}
D_M(E) &=& \frac{1}{\alpha}\ln\left(\frac{1+\sqrt{1+E/U_0}}{1-\sqrt{1+E/U_0}}\right) \cr
&=&
\frac{1}{\alpha}\ln\left(\frac{(1+\sqrt{1+E/U_0})^2}{-E/U_0}\right) \cr
&=&
\frac{2}{\alpha}\ln\left(\sqrt{-\frac{U_0}{E}}+\sqrt{-\frac{U_0}{E}-1}\right)\, .
\end{eqnarray}

Let us now do the same thing for the P\"oschl-Teller potential. An analogous procedure to determine the roots of the equation $E = U_{PT}(x)$ yields
\begin{equation}
\label{eq:turningPT}
x_{\pm}=\pm\frac{1}{\alpha}\operatorname{arcosh}(\sqrt{-U_0/E})\, .
\end{equation}
Therefore, the distance between these two turning points is given by 
\begin{equation}
\label{eq:DPT}
D_{PT}(E) = x_+ - x_- = (2/\alpha)\operatorname{arcosh}(\sqrt{-U_0/E})\, .
\end{equation}
Naively, equations (\ref{eq:diamMorse}) and (\ref{eq:DPT}) do not seem to coincide. However, after applying the mathematical identity 
$\operatorname{arcosh}(x)=\ln(x+\sqrt{x^2-1})$
to $D_{PT}(E)$, it becomes evident that $D_M(E) = D_{PT}(E)$. This means, as we had anticipated, that these two potentials are sheared relative to each other for $-U_0 < E <0$. 

This means that, in order to obtain the periods of oscillations for any of these two potentials (for $-U_0 < E <0$), we may choose whichever we want, as they will be the same for both potentials. Let us do this calculation explicitly for the Morse potential. Using Eq. (\ref{eq:periodo-pot}) and making the same change of variables as that used to obtain Eq. (\ref{eq:changevarMorse}), we get 
\begin{align}
	\label{eq:periodoMorse}
	\tau_M(E)&= \sqrt{2m} \int_{x_-}^{x_+} \frac{dx}{\sqrt{E-U_0(e^{-2\alpha x} - 2e^{-\alpha x})}} \nonumber \\
	&= \sqrt{\frac{2m}{U_0\alpha^2}} \int_{u_-}^{u_+} \frac{du}{\sqrt{\frac{E}{U_0} u^2+2u+1}}\, .
\end{align}
The new change of variables $w=(u \sqrt{ - E/U_0} + \sqrt{-U_0/E})$ leads to
\begin{align}
	\label{eq:periodMorse2}
	\tau_M(E) &= \sqrt{\frac{2m}{U_0\alpha^2}} \int_{w_-}^{w_+} \frac{dw}{\sqrt{w^2-(\frac{U_0}{E}+1)}} \nonumber \\
	&= \frac{\pi}{\alpha} \sqrt{\frac{2m}{E}}\, .
\end{align}
We leave for the interested reader the task of checking explicitly through an analogous calculation that the periods associated to the P\"oschl-Teller potential (also for 
$-U_0 < E < 0$) are indeed the same as those found for the Morse potential written in the previous equation.

\subsection{General prescription for sheared potentials}

\label{sec:prescription}

After these examples, one might ask what is the general prescription for constructing sheared potential wells. We may always begin with the single symmetric potential well $U(x)$, associated to a given period function, so that $x_R(U) = - x_L(U)$. Then, in order to shear this potential we just displace the function horizontally by $\delta (U)$ to obtain new functions $\tilde{x}_R(U)$ and $\tilde{x}_L(U)$ respectively for the right-side and the left-side branches\cite{Osypowski-1986,Anton-parabolashear2008}:
\begin{equation}
\label{eq:sheardelta}
\left\{ \begin{array}{lr} \tilde{x}_R(U)&=x_R(U)+\delta(U)\\
\tilde{x}_L(U)&=x_L(U)+\delta(U) \end{array}\right.
\end{equation}
Consequently, the difference between the corresponding turning points is preserved for each value of $U$, namely, $D(U)=\tilde{x}_R(U)-\tilde{x}_L(U)=x_R(U)-x_L(U)$. 
Both $\tilde{x}_R$ and $\tilde{x}_L$ must be single-valued. If this condition is met, taking the inverse piecewise function gives a new potential $\tilde{U}(x)$ which is sheared from $U(x)$.

\section{Shearing tracks}
\label{sec:generalization}

Though not so popularized, sheared potential wells are very well established. Inspired by the surprising properties of these potential wells, we now turn our attention to a different problem which, as far as the authors know, had not yet been investigated. Instead of the one-dimensional motion driven by a given potential well $U(x)$, we consider the two-dimensional movement of a particle sliding along a frictionless track under constant gravity. As constructing tracks is generally more feasible than supplying the conditions for obtaining an arbitrary potential well, this technique may prove useful for actual visualization of the motions under study. 

We state the problem as follows: Let a particle be subject solely to an uniform gravitational field, so that the gravitational potential energy is $U(y)=mgy$, and be constrained to move along a smooth track contained in the ${\cal O}xy$ plane. Due to this constraint, the motion has only one degree of freedom. The shape of the track, described by the function $f: x \longmapsto y = f(x)$, determines the net force on the particle at each point. For convenience, we choose a coordinate system so that the single minimum of the track coincides with the origin of the axes. Though the shape of the track uniquely determines the periods of the motion as a function of its energy, $\tau(E)$, it is not obvious whether the knowledge of $\tau(E)$ uniquely determines the shape of the track along which the motion takes place. In other words, it is natural to ask what tracks give rise to oscillations with a given period function $\tau: E \longmapsto\tau(E)$.

As it will become evident further, it is convenient to use the arc-length coordinate $s$ to parametrize the track instead of the cartesian coordinate $x$, and henceforth we denote the shape of the track by the function $y: s \longmapsto y(s)$, with $s=0$ at the origin of the ${\cal O}x$ and ${\cal O}y$ axes. Setting the maximum height $H$ achieved by the particle also sets the mechanical energy $E=mgH$ of the system, and the turning points $s_1$ and $s_2$ are just given by the roots of the equation $y(s)=H$. From the conservation of mechanical energy, it is straightforward to obtain the corresponding period of oscillations $\tau(H)$ for an arbitrary $H$: this period is simply twice the time spent by the particle to move from $s_1$ to $s_2$ along the track. Denoting by $v$ the scalar velocity of the particle at position $s$ along the track, we have
\begin{equation}
\tau(H) = 2 \int_{s_1}^{s_2}\! \frac{ds}{v} = \sqrt{\frac{2}{g}}\int_{s_1}^{s_2}\frac{ds}{\sqrt{H-y(s)}}\, .
\end{equation}
Note that, as $s_1$ and $s_2$ depend on $H$, $\tau$ is a function only of $H$, uniquely determined by knowledge of the track $y(s)$. Furthermore, the previous equation neatly displays only variables with a clear geometrical meaning.

We now wish to tackle the inverse problem, which is to find the $y(s)$ that constrains the motion to a given periodic behavior $\tau(H)$. We apply the same mathematical procedure as Landau and Lifshitz to this new case. We invert the function $y(s)$ by means of the piecewise function split in $s_R(y)$ and $s_L(y)$ respectively for the right and left branches of the track, and integrate using a trick analogous of that presented in Section \ref{sec:shearedreview}, to give:
\begin{equation}
\label{eq:length-sheared}
s_R(y)-s_L(y)=\frac{1}{\pi}\sqrt{\frac{g}{2}}\int_0^y \frac{\tau(H)dH}{\sqrt{y-H}}.
\end{equation}
The reader may note that this result is akin to Eq. (\ref{eq:xRxL}), with $x \rightarrow s$, $U \rightarrow y$ and $E \rightarrow H$, but we emphasize that it has a different meaning and interpretation.

The previous equation shows that knowledge of the period function $\tau(H)$ does not uniquely determine the shape of a track, but rather its length $L(y)=s_R(y)-s_L(y)$ below every height $y>0$. Two different tracks $y(s)$ and $\tilde y(s)$ will lead to motions with the same period function $\tau (H)$ if $s_R(y) - s_L(y)=\tilde{s}_R(y) - \tilde{s}_L(y)$, for every $y$. 
We call these tracks {\it length-sheared} relative to one another in analogy with sheared potentials. 

This means that, concretely, we may construct a track made of measuring tape against a wall with a graph paper pattern, and measure the period of oscillations of a small marble moving on it. Then we slide the tape carefully as to ensure that the length of the tape under every horizontal line remains the same. This transformation will generate a new track, related to the first by length-shearing, and oscillations along this second track will have the same period as the first one for every maximum height $H$. This is noteworthy, since a similar practical procedure is not generally possible with sheared potentials.

A comment is in order here. Suppose we consider only the motions of a particle which moves down a frictionless track characterized by a function $f: x \longmapsto y = f(x)$ and denote by $\tau(H)$ the time spent by the particle to reach the origin at $y=0$ once abandoned at rest at an arbitrary height $y = H$. Again, given the track, $\tau(H)$ is uniquely determined. However, in this case, the knowledge of $\tau(H)$ would also uniquely determine the shape of the track (note that we are not talking about periodic motions but only downward motions). This problem was first solved by Niels Abel in 1826 \cite{Abel-1826}. In 2010, Mu\~noz and Fern\'andez-Anaya \cite{Munoz-2010} discussed Abel's result for particular curves for which the corresponding periods are proportional to a fractional power of $H$. Also, the same authors ({\it et al.}) present an illuminating set of direct and inverse problems involving beads on a frictionless rigid wire in a paper from 2011\cite{Munoz-2011}.

\section{Round-trip tautochrones}
\label{sec:tautochrone}

\subsection{Brief historical survey}

Before we apply our previous result to the tautochrone problem it is worth saying a few words about the tautochrone curve which played an important role in the history of classical mechanics of the 17th century. At that time, measuring the latitude was very simple but measuring the longitude, which was literally of vital importance for the sea navigations, was far from being an easy task, since it demanded a quite accurate measurement of time \cite{Longitude}. The pendulum clock, constructed by the great dutch physicist, mathematician and astronomer Christiaan Huygens in 1658 (Galileo had already tried to construct a pendulum clock but he never finished and the patent of this invent was given to Huygens) improved at least in one order of magnitude the accuracy of time measurements, but this was not enough to guarantee a safe measurement of the longitude. 

With the purpose of improving maritime chronometers, Huygens started to look for an isochronous pendulum, since he knew that the simple pendulum was isochronous only for small amplitudes. A maritime chronometer constructed with an isochronous pendulum would not change the period of its oscillations even if the corresponding amplitudes changed due to a rough sea. Huygens knew that if he put lateral obstacles of appropriate shape near a simple pendulum he could achieve his purpose but, unfortunately, he was not able to find empirically the exact shape of such lateral obstacles. 

Then destiny came to his aid. Blaise Pascal, the famous french physicist, mathematician and philosopher, who had abandoned science after a religious epiphany in 1654, had an unbearable toothache in 1658 that seemed to resist any alleviating efforts. In a desperate attempt, Pascal decided to think of mathematics, particularly, in some problems on the cycloid that the french abbey Mersenne had passed to him. Coincidentally or not, the pain disappeared completely and Pascal interpreted this fact as a divine sign for him to go on thinking of problems involving this curve. And so he did. He solved many of Mersenne's problems and formulated a few more. However, instead of publishing them, he decided to propose a contest composed of six problems involving the cycloid. 

Many important scientists of that time were encouraged somehow to participate in that contest, including Huygens. Once he had become an expert on the cycloid, Huygens decided to check if, by any chance, the cycloid would solve his problem. Fortunately, Huygens found that the cycloid was a tautochrone (a curve over which a particle would slide without friction under the action of a constant gravity with periods independent of the height where it was abandoned). But he still needed to find out the shape of the lateral obstacles that would make a pendulum describe a cycloidal trajectory. In other words, he had to find out what we call the evolute of the cycloid. Again, he tried the cycloid and once more he was successful: the evolute of the cycloid is the cycloid (shifted and out of phase). He was very lucky, since it is not very common that a curve is the evolute of itself. More details on this history is nicely told in Gindikin's book \cite{Tales}.
 
Among other things, Huygens worked on trying to improve clocks for almost four decades, but his cycloidal pendulum clock, as well as his isochronous conical pendulum clock, did not succeed as maritime chronometers. However, his legacy on developments of curves, evolutes and involutes which had their origins in his study of clocks can be sensed until nowadays in many different areas from differential geometry to quasicrystals \cite{Arnold}. 

\subsection{Obtaining the tautochrone curve}

Eq. (\ref{eq:length-sheared}) teaches us how to construct tracks on which particles oscillate with a given period $\tau(H)$. If we impose the track to be symmetric, then it is uniquely determined. We shall now apply the previously discussed techniques to obtain the tracks on which the period does not depend on the energy, that is, the solutions to the tautochrone problem. As these tracks must have upward and downward branches allowing for periodic motion, we call them {\it round-trip tautochrones}.

Choosing a suitable energy-independent period function $\tau(E)=\sqrt{\kappa}$, where $\kappa$ is a positive constant with dimensions of time squared, and demanding that the track be symmetric allows us to compute Eq. (\ref{eq:length-sheared}) exactly, as the integral becomes elementary:
\begin{align}
\label{eq:cycloidparam}
s(y)&=\frac{1}{2\pi}\sqrt{\frac{g\kappa}{2}}\int_0^y\frac{dH}{\sqrt{y-H}} \nonumber \\
&= \frac{1}{2\pi}\sqrt{\frac{g \kappa}{2}} (2\sqrt{y}) \Rightarrow \nonumber \\
\Rightarrow y &= \frac{2\pi^2}{\kappa g} s^2 \, .
\end{align}
The above result is in perfect analogy with the harmonic oscillator. As expected, it also corresponds to the arc-length parametrization of a cycloid whose generating circle has radius $r = \kappa g/(4\pi)^2$, $y = (1/8r)s^2$. The period of oscillations in terms of $r$ and $g$ is readily given by
\begin{equation}
\label{eq:periodsymcycloid}
\tau(r)=4\pi\sqrt{\frac{r}{g}} \, .
\end{equation}
Although the cycloid is indeed the single symmetric round-trip tautochrone, we may length-shear it in infinitely many ways. 
We will now present unconventional solutions to the tautochrone problem, given by asymmetric tracks. 

\subsection{Half-cycloid branches}

Let us begin with a symmetric cycloid whose generating circle has radius $r$. We first attempt an asymmetric length-sheared solution using cycloids with different generating radii. 
We consider for the left-side and right-side branches, respectively $\tilde{r}_L$ and $\tilde{r}_R$, in analogy to what we have presented in section \ref{ssec:powerlaw} for coefficients of power-law potentials.

\begin{figure}[!hbt]
\begin{center}
\includegraphics[width=\linewidth]{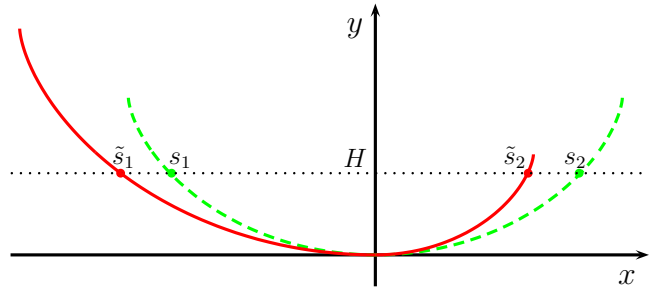}
\caption{Symmetric cycloid (dashed green line) and an isochronous asymmetric track made from different cycloids (solid red line).}
\label{fig:rightleft_cycloid1}
\end{center}
\end{figure}

For both tracks to give rise to oscillations of equal periods, they must be length-sheared, so the total length of the track must be the same below every height $y$. The relation between the radii follows directly from applying constraint
\begin{equation}
\label{eq:cond_cycloidlength1}
s_R(y)-s_L(y) = \tilde{s}_R(y)-\tilde{s}_L(y)\, .
\end{equation}
From the relation $y = (1/8r)s^2$, it follows immediately that
\begin{equation}
\label{eq:cond_cycloidlength2}
2\sqrt{r}=\sqrt{\tilde{r}_R}+\sqrt{\tilde{r}_L} \, .
\end{equation}
We present one such solution in Fig. \ref{fig:rightleft_cycloid1}.

Alternatively, we may obtain such a curve by imposing that the periods of oscillations along both the symmetric and the asymmetric tracks be the same (provided the maximum height of the motion is contained in the lower branch of the curve). The interested reader may easily verify, by means of Eq. (\ref{eq:periodsymcycloid}), that the condition obtained is the same as in Eq. (\ref{eq:cond_cycloidlength2}).

\subsection{Completing tautochrones}
Eq. (\ref{eq:length-sheared}) allows for a very broad range of solutions for round-trip tautochrones which are not restricted to branches of cycloids. We may obtain such a track using a prescription similar to the one presented in section \ref{sec:prescription}, 
namely, by applying (to the original cycloidal track) equations analogous to those written in Eq. (\ref{eq:sheardelta}), but now with a length-shearing function $\delta(y)$, instead of the shearing function $\delta(U)$. With this procedure in mind, if we choose an arbitrary function for the left branch of the track, say $\tilde{y}_L(s)$, we will obtain the corresponding $\tilde{y}_R$ function for the right branch that completes the tautochrone track.

Since any track constructed in this way will be length-sheared from the original cycloidal track (which corresponds to the single symmetric tautochrone), we see that this procedure provides a method for generating as many tautochrone tracks as we want. All tracks constructed in this way will share the property of isochronous motions, that is, all of them will lead to periodic motions with height-independent periods.

\begin{figure}[!hbt]
\begin{center}
\includegraphics[width=\linewidth]{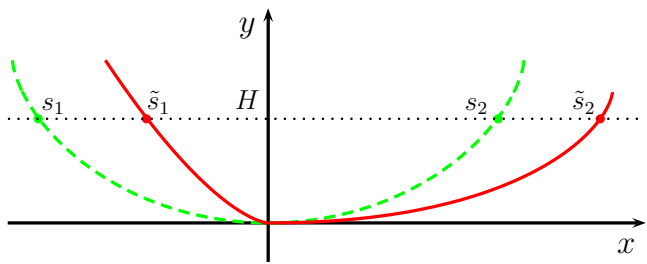}
\caption{Cycloid (dashed green line) and an isoperiodical asymmetric tautochrone having a semicubical parabola for its left branch (solid red line). In this example, we have set $r=1/8$ and $\alpha=2$.}
\label{fig:rightleft_cycloid2}
\end{center}
\end{figure}

An explicit example is appropriate here. Let us set the left branch of the track to the semi-cubical parabola $\tilde{y}_L=\alpha(-x)^{\frac{3}{2}}$ ($\alpha>0$). We compute the arc-length in order to parametrize it conveniently:
\begin{equation}
\label{eq:semicubica_s}
\tilde{s}_L(y)=\frac{8}{27\alpha^2}\left[ \left(\frac{9\alpha^\frac{4}{3}}{4}y^{\frac{2}{3}}+1 \right)^{\frac{3}{2}}-1 \right] \, .
\end{equation}
We may set the period of oscillations by adjusting the radius $r$ of the original cycloid $8ry=s^2$, in accordance to Eq. (\ref{eq:periodsymcycloid}). So by imposing that the new track satisfies the length-shearing relation to this cycloid, the right-side branch of the track is given by the equation:
\begin{equation}
\label{eq:semicubica_delta}
\tilde{s}_R(y) = 2\sqrt{8ry}  + \tilde{s}_L(y) \, .
\end{equation}
A solution to the track having a semi-cubical parabola for its left branch is shown in Fig. \ref{fig:rightleft_cycloid2}.

Since we chose an arbitrary function for the left branch of the tautochrone, the time it spends on each branch will depend on the maximum height. However, the geometrical condition given by length-shearing guarantees that the difference of time the moving particle spends on the left branch (with respect to the cycloid) will be compensated exactly on the right branch. 

\section{Final remarks}
\label{sec:remarks}

In the scope of one-dimensional systems in classical mechanics, we have reviewed the result that oscillations with a given period as a function of the energy correspond not to one, but to an infinite family of potential wells, which satisfy the shearing condition. Particularly, isochronous oscillations are not a property associated uniquely to the quadratic potential well (harmonic potential) but to any potential well which is sheared from it. 

Then we have extended this result to the motions of a particle moving along frictionless tracks contained in a vertical plane and subject to a constant gravity. We have shown that if a new condition, called length-shearing, is imposed to the tracks, their corresponding periods, as functions of the maximum height achieved by the particle, will be the same. 

Applying the length-shearing condition to a cycloid, the original tautochrone discovered by Huygens in the 17th century, we have shown that it is possible to obtain an infinite family of solutions to the tautochrone problem.

The actual construction of tracks --- which can be done with a 3-D printer, for instance --- is far more practical than providing the conditions for a certain potential energy $U(x)$ to drive the motion in a given period. Also, it is worth mentioning that the techniques employed in this work are accessible to undergraduate students. Hence the information presented here can be useful for enriching courses on both theoretical and experimental university-level classical mechanics.

Finally, for pedagogical reasons, we would like to outline the most important aspects of the present work: 

\begin{itemize}
	\item The study of sheared potentials, although being a well established problem, still holds a few surprises, such as the shearing relation between the Morse and P\"oschl-Teller one dimensional potentials.
	\item There is a new condition analogous to shearing in one-dimensional potential wells, called length-shearing, which can be applied to frictionless tracks contained in vertical planes.
	\item There is not one, but infinite tautochrones.
	\item We can make a tautochrone by choosing any monotonic function $y: x \longmapsto y(x)$ for the shape of one of its branches, and then computing the corresponding complementary branch simply by imposing that they are length-sheared to the cycloid.
	
\end{itemize}

We leave for the interested reader the task of exploring new examples of tracks exhibiting different properties regarding the periods of their respective driven motions, further applying length-shearing.

The authors thank Felipe Rosa for his very fruitful insights. CF is also indebted to some members of the Theoretical Physics Department of University of Zaragoza, particularly, to M. Asorey, J. Esteves, F. Falceto, L.J. Boya and J. Cari\~nena for enlightening discussions. This work was partially supported by the brazilian agencies CNPq and FAPERJ.

\end{document}